\documentclass[12pt]{article}

\usepackage[body={17.5cm, 23cm},right=2cm]{geometry}

\usepackage{booktabs}
\usepackage{caption}

\usepackage{color}
\usepackage{graphicx}
\usepackage{epsf}
\usepackage{graphicx,epsfig}
\usepackage{multirow}

\usepackage{amsmath, amsthm, amssymb}

\usepackage{epsfig}
\usepackage{cite}
\usepackage{color,colordvi}

\newcommand{\be}{\begin{eqnarray}}
\newcommand{\ee}{\end{eqnarray}}
\newcommand{\bi}{\begin{itemize}}
\newcommand{\ei}{\end{itemize}}

\def\<{\left\langle}
\def\>{\right\rangle}

\def\s{\sigma}
\def\nn{\nonumber}
\def\p{\partial}
\def\ls{\left[}
\def\rs{\right]}
\def\lc{\left\{}
\def\rc{\right\}}

\newcounter{hran}

\makeatletter
\renewcommand\section{\@startsection {section}{1}{\z@}%
                               {-3.5ex \@plus -1ex \@minus -.2ex}%
                               {2.3ex \@plus.2ex}%
                               {\normalfont\large\bfseries}}
\makeatother

\begin{document}

\vspace{0.5cm}

{\ }

\vspace{0.8cm}

\begin{center}

\def\thefootnote{\fnsymbol{footnote}}

{\large \bf 
  Non-minimal derivative couplings and inflation in supergravity 
  %
}
\\[1.5cm]
{\ }
\\[0.3cm]
{\large Ioannis Dalianis$^{(a)}$ and  Fotis Farakos$^{(b)}$}
\\[0.2cm]

\vspace{.3cm}
{\normalsize {\it $^{(a)}$  Physics Division, National Technical University of Athens, \\ 15780 Zografou Campus, Athens, Greece}}

\vspace{.3cm}
{\normalsize {\it $^{(b)}$  Institute for Theoretical Physics, Masaryk University, \\611 37 Brno, Czech Republic}}

\vspace{.3cm}
{\normalsize  E-mail: dalianis@mail.ntua.gr, fotisf@mail.muni.cz }


\end{center}

\vspace{2cm}

\hrule \vspace{0.3cm}
\small  \noindent \textbf{Abstract} 
\noindent

In this article we motivate and review the embedding of the gravitationally enhanced friction mechanism in supergravity. 
The very interesting feature is that inflationary models which utilize this mechanism 
drive inflation for a wider range of parameter values and  
predict lower values for the tensor-to-scalar ratio.

\vspace{0.5cm}  \hrule
\vskip 2.0in

\noindent
{\it Proceedings of the Corfu Summer Institute 2013 \\
 		 August 31 - September 11, 2013\\ 
 		 Corfu, Greece}

\def\thefootnote{\arabic{footnote}}
\setcounter{footnote}{0}


\newpage



\baselineskip= 19pt

\vspace{5mm}

\def\thefootnote{\arabic{footnote}}
\setcounter{footnote}{0}

\def\ls{\left[}
\def\rs{\right]}
\def\lc{\left\{}
\def\rc{\right\}}

\def\p{\partial}

\def\S{\Sigma}

\def\nn{\nonumber}

\section{Gravitationally enhanced friction and inflation}

The inflationary paradigm \cite{Lyth:1998xn,Liddle:2000cg,Mukhanov:2005sc}  is strongly supported by the observational data \cite{Ade:2013uln,Ade:2015lrj}. 
The usual challenge one faces in embedding inflation in a microscopic theory 
is the construction of the appropriate scalar potential. 
Restricted by our little knowledge on the exact description of the theory at the energy scales of inflation, 
one uses an effective approach and various symmetries, naturalness arguments or other mechanisms are invoked to 
control the effective form of the inflationary potential.  
The problem of maintaining the form of the potential during the trans-Planckian excursions of the inflaton field 
becomes even more severe in supergravity due to the restriction placed to the potential by supersymmetry 
and also due to the generic cut-off of the theory being the Planck scale ($M_P$) itself.

Among the proposals of how to achieve a generic inflationary phase  
is the gravitationally enhanced friction mechanism (GEF) 
\cite{Germani:2010gm,Germani:2010hd,Germani:2011mx,Germani:2011bc,Dalianis:2014sqa,Dalianis:2014nwa}. 
This mechanism is generated by the following Lagrangian 
\be \label{GEF}
{\cal L}= \frac{M_P^2}{2} \sqrt{-g} R 
-\frac{1}{2} \sqrt{-g} \left(g^{mn} -M^{-2}_*G^{mn} \right)\partial_m\phi\partial_n\phi
-\sqrt{-g} V(\phi),  
\ee
where $\phi$ is the inflaton and $G^{mn}=R^{mn}-\frac12 g^{mn}R$ is the Einstein tensor. 
This kinematic coupling of a scalar field to gravity does not give rise to ghost instabilities, 
and belongs to a more generic class of theories which can be also used for cosmology 
\cite{Horndeski:1974wa,Amendola:1993uh,Capozziello:1999xt,Sushkov:2009hk,Germani:2010ux,
Saridakis:2010mf,Germani:2011ua,Charmousis:2011bf,
Charmousis:2011ea,Koutsoumbas:2013boa,Kolyvaris:2011fk,Dent:2013awa,Minamitsuji:2015nca,Yang:2015pga}. 
We assume here a full polynomial potential without symmetry suppressed terms, 
representative of a generic supergravity setup 
\be
V(\phi)=\sum_n\lambda_nM^{4-n}_P\phi^n.
\ee 
In the large field models of inflation the inflaton field has a value larger than the Planck mass, $M_{P}$. 
This general potential cannot serve as large field inflationary model because the non-renormalizable terms, if not suppressed, spoil the flatness of the potential. 

The non-minimal coupling of the kinetic term of the scalar field with the Einstein tensor $G_{mn}$ 
during a de Sitter phase takes the simple form 
\be
M^{-2}_*G^{mn}=-3M^{-2}_*H^2 g^{mn}.
\ee 
For $H M^{-1}_*\gg 1$ the kinetic coupling implies that the canonically normalized scalar field is the $\tilde{\phi}= \sqrt{3}H M^{-1}_* \phi$. 
This rescaling recasts the polynomial potential 
in terms of the canonically normalized inflaton $\tilde{\phi}$ to the form
\begin{equation}
V(\tilde{\phi})=\sum_n \lambda_n M^{4-n}_P \left(\frac{\tilde{\phi}}{\sqrt{3}HM^{-1}_*}\right)^n\,.
\end{equation}
The non-renormalizable terms $\sum^\infty_{n=4} \lambda_n M_{P}^{4} \left({\tilde{\phi}}\times({\sqrt{3}H M^{-1}_* M_{P}})^{-1}\right)^n $ are suppressed by the ``enhanced'' mass scale $\sqrt{3}H M^{-1}_* M_{P}$. The slow roll parameters require $\tilde{\phi}>M_{P}$ and, hence, these higher order terms can be neglected and sufficient inflation can take place given that 
\begin{equation} \label{fv1}
M_{P} < \tilde{\phi} \ll M_{P}(H M^{-1}_*)\,.
\end{equation}
In terms of the field $\phi$, which has non-canonical kinetic term, the above field-space region translates into
\begin{equation} \label{fv2}
\frac{M_{P}}{H M^{-1}_*} < {\phi} \ll M_{P}\,.
\end{equation}
This finding is of central importance since we will work in a supersymmetric context. 

We consider our theory as an effective one valid below some ultra-violet cut-off that we generally identify with the $M_P$. The field-space region (\ref{fv2}) allows inflation to be realized for a generic form of potentials and reliable conclusions in this context can be derived. It can be said that the kinetic coupling theory is tailor-made for realizing an inflationary phase. 
From a different perspecive, if there is an internal symmetry that forbids the non-renormalizable terms and thereby suppresses the coefficients $\lambda_n$ for $n\geq 5$ then inflation can be implemented in a much larger  field-space region, than in the conventional (GR limit) large field inflationary models, that reads:
$\phi>M_P/(HM^{-1}_*)$ .

The plateau like potentials are not common in minimally coupled theories. It may indicate that the plateau-like behaviour is result the of nonminimal dynamics such as the derivative couplings to the Einstein tensor demonstrate. For example steep potentials for fields $\phi$ with nonminimal kinetic coupling can yield potentials of moderate steepness for a minimally coupled $\varphi$ that can well fit the data. There exists the particularly interesting mapping \cite{Germani:2011mx,Dalianis:2014nwa} 
\begin{equation}
\phi^n \quad \longleftrightarrow \quad \varphi^\frac{2n}{n+2} . 
\end{equation}
The crucial implication of the non-minimal derivative coupling is that generic and steep potentials can inflate the Universe.  Otherwise, one has to impose special symmetries or truncate higher order terms in order to preserve the flatness of the potential. If the nonminimal derivative coupling (\ref{GEF}) is present then a steep potential  e.g. the $\phi^{100}$ appears like $\varphi^2$ and the potential  $\phi^2$ appears like  $\varphi$.

Finally, we comment on another interesting feature of these kind of higher derivative theories. Generally, the plateau inflationary potentials suffer from the initial conditions problem \cite{Dalianis:2015fpa}. The effective potentials that originate from the nonminimal derivative coupling theories (\ref{GEF}) can be flat enough to well fit the data \cite{Ade:2015lrj}. However, they account for large field models and  inflation can start from Planck energy densities requiring the minimum homogeneity for the accelerating expansion to start.

\section{Non-minimal derivative couplings in supergravity and D-term potential}

The non-minimal derivative coupling can be successfully embedded in the  new-minimal supergravity. The new-minimal supergravity \cite{Sohnius:1981tp,Ferrara:1988qxa} is the locally supersymmetric theory of the gravitational multiplet
\begin{equation}
e^a_m\, ,~~\psi_m^\alpha \,  , ~~{\cal A}_m\, , ~~ B_{mn}\ .
\end{equation}
The first two fields are the vierbein and its superpartner the gravitino, 
a spin-$\frac{3}{2}$ Rarita-Schwinger field. 
The last two fields are auxiliaries. 
The real auxiliary vector ${\cal A}_m$ gauges the $U(1)_\text{R}$  symmetry. 
The auxiliary $B_{mn}$ is a real two-form appearing only through its dual field strength $H_m$, 
which satisfies $\hat D^a H_a =0$ for the supercovariant derivative $\hat D^a$.

We will employ  the new-minimal supergravity  superspace \cite{Ferrara:1988qxa}.  
The  free Lagrangian is given by
\begin{equation}
\label{sugra} 
{\cal L}_{\text{new-min}}= - 2 M^2_P  \int d^4 \theta E V_{\text{R}} . 
\end{equation}
Here $V_{\text{R}}$ is the gauge multiplet of the R-symmetry, 
which (in the appropriate WZ gauge) contains the auxiliary fields in its vector component $- \frac{1}{2} [\nabla_\alpha , \bar \nabla_{\dot \alpha} ]  V_{\text{R}} | = {\cal A}^-_{\alpha \dot \alpha} 
= {\cal A}_{\alpha \dot \alpha} - 3 H_{\alpha \dot \alpha} $  
and the Ricci scalar in its highest component  
$ \frac{1}{8} \nabla^\alpha \bar \nabla^2 \nabla_\alpha  V_{\text{R}} |= - \frac{1}{2} \left( R + 6 H^a H_a \right) $. 
The symbol $E$  stands for the super-determinant of new-minimal supergravity.  
The bosonic sector of Lagrangian (\ref{sugra}) is
\begin{equation}
\label{freesugra}
{\cal L}_{\text{new-min}}= 
  M^2_P \, e \, \left( \frac{1}{2}
R + 2{\cal A}_a H^a-3H_a H^a\right) .
\end{equation}

For a  chiral superfield with vanishing R-charge the kinematic Lagrangian reads 
\be
{\cal L}_0 = \int d^4  \theta \, E  \bar \Phi \Phi ,  
\ee
the bosonic sector of which is 
\be
{\cal L}_0 = A\Box \bar A+  F\bar F 
-i H^m \left(A\,\partial_m \bar A- \bar A \partial_m A\right) . 
\ee
Here we use the definitions for the component fields $\Phi|=A$ and $-\frac14 \nabla^2 \Phi|=F$.

The superspace Lagrangian that gives rise to the non-minimal derivative coupling in the bosonic sector is \cite{Farakos:2012je} 
\be
\label{NMDC}
{\cal L}_{M_*} =  i M^{-2}_* \int d^4 \theta \, E \, [ \bar \Phi E^a \nabla_a \Phi ] + c.c.  , 
\ee
where $E^a$ is a curvature real linear superfield  of the new-minimal supergravity 
which satisfies the superspace Bianchi identities 
\be
\begin{split}
\nabla^2 E_a &=0  , 
\\
 \bar \nabla^2 E_a &=0 , 
\\
\nabla^a E_a &=0 .   
\end{split}
\ee
The $E^a$ superfield has bosonic components  
\be
E_a | = H_a , 
\ee
and 
\be
\frac{1}{4}  \bar \s_a^{\dot \alpha \alpha}  [\nabla_\alpha , \bar \nabla_{\dot \alpha} ] E_b | = 
\frac{1}{2} ( G_{ab} - g_{ab} H^c H_c - 2 H_a H_b - {}^* {\cal F}_{ab} ) , 
\ee
where $G_{mn} = R_{mn} - \frac{1}{2} g_{mn} R$ is the Einstein tensor and 
$ {\cal F}_{mn} = \p_m {\cal A}_n - \p_n {\cal A}_m $ is the field strength of the supergravity auxiliary field ${\cal A}_m$. 
For a discussion and derivation of the Lagrangian (\ref{NMDC}) see \cite{Farakos:2012je}. 
Moreover this term is not known how to be constructed in the old-minimal supergravity formulation \cite{Wess:1992cp}. 
Discussion on more theoretical aspects  of the embedding of this and similar couplings 
in new-minimal supergravity can be found in \cite{Assel:2014tba}. 
The bosonic sector of  Lagrangian (\ref{NMDC}) is 
\begin{equation}
\begin{split} 
{\cal L}_{M_*} = & M^{-2}_* \left[ 
\, G^{ab}\partial_b \bar A \,\partial_a A
+2F \bar F H^a {\cal A}_a
-2F \bar F H^aH_a\right.
\\
\label{non-min}
&
+i  H^a \left(\bar F \partial_a F-F\partial_a \bar F\right) 
- \partial_bA\,\partial^b \bar A H_aH^a \\
& \left. +2\,H^a\partial_aA\,H^b\partial_b \bar A
-i H_c \left(\partial_b \bar A \,{\cal{D}}^c\partial^b A 
-\partial_b A\,{\cal{D}}^c\partial^b \bar A \right)\right] . 
\end{split}
\end{equation}
Note that this term, although it contains higher derivatives, 
does not lead to ghost states or instabilities. 
The ghost instabilities are in fact evaded due to the vanishing R charge of the chiral superfield $\Phi$.  
On the other hand, 
the vanishing R-charge forbids the self-coupling via a superpotential due to the R-symmetry. 
In other words, this superfield is not allowed to have a superpotential. 
Moreover it is also not allowed to be gauged, 
since this will also give rise to ghost instabilities 
via inconsistent derivative couplings of the gauge fields to curvature. 
The only remaining option 
is the indirect introduction of self-interaction via a   gauge kinetic function \cite{Dalianis:2014sqa}.

Theories with a non-minimal derivative coupling  can lead to galileons  in specific decoupling limits 
for gravitation \cite{Germani:2012qm}. 
This has been also done in a supersymmetric setup. 
Taking the limit $M_P \rightarrow \infty$ but keeping  $M_*^2 M_P$ fixed one can find the 
supersymmetrization of the quartic complex galileon \cite{Farakos:2013fne}. 
In particular it has been found that these theories have a new superspace symmetry which is the supersymmetric generalization of the galileon symmetry 
\be
\label{sg}
\Phi \rightarrow \Phi + a + b^m y_m , 
\ee
where $a$ is a complex constant, $b_m$ a complex constant vector, and $y_m = x_m + i \theta \s_m \bar \theta$. 
The superspace shift (\ref{sg}) reduces to the standard galileon shift for the complex scalar $A$ in the lowest component.

Turning back to inflation, we still have to introduce a scalar potential. 
It has been shown in \cite{Dalianis:2014sqa} that a scalar potential can be consistently generated for $A$ 
if we utilize a Fayet-Iliopoulos term. 
In particular consider the gauge sector 
\be
\label{GGG}
{\cal L}_g = \frac{1}{4} \int d^2 \theta {\cal E} f(\Phi) W^2 (V) + c.c. + 2 \xi \int d^4 \theta E V , 
\ee
with $W_{\alpha} (V) = - \frac{1}{4} \bar \nabla^2 \nabla_\alpha V $ 
where $f(\Phi)$ is a holomorphic function of the chiral superfield $\Phi$ 
and $\xi$ is the Fayet-Iliopoulos parameter of mass dimension two. 
The coupling (\ref{GGG}) leads to a scalar potential of the form 
\be
\label{VA1}
{\cal V} = \frac{1}{2} \frac{\xi^2}{\text{Re}f(A)}.
\ee
This is a D-term potential \cite{Lyth:1998xn,Binetruy:1996xj}.

To wrap it up, 
the theory under consideration is given by \cite{Dalianis:2014sqa} 
\be
\label{total}
{\cal{L}}_{\text{total}}
= {\cal{L}}_{\text{new-min}}
+ {\cal L}_{\text{g}}
+ {\cal L}_0
+ {\cal L}_{M_*} , 
\ee
which after we integrate out the auxiliary fields has bosonic sector 
\be
\label{total-onshell}
\begin{split}
e^{-1}{\cal{L}}_{\text{total}} =& \frac{M^2_P}{2} R  + A\Box
\bar A 
+M^{-2}_* 
\, G^{ab}\,\partial_a
\bar A\, \partial_b A\ -\frac{1}{2} \frac{\xi^2}{\text{Re}f(A)} 
\\
& -\frac{1}{4}  \text{Re}f(A) F^{mn} F_{mn} 
+ \frac{1}{4} \text{Im}f(A) F^{mn} \ ^*F_{mn}.
\end{split}
\ee

Even though we suggest a D-term inflation without a superpotential the generation of the inflationary potential is in principle not protected by any symmetry and in the most general case we cannot forbid the higher order terms. 
A resolution to the $\eta$-problem in  supergravity  can be given by a symmetry that suppresses the F-term part of the scalar potential. In the presence of such a symmetry the potential is naturally dominated by a Fayet-Iliopoulos D-term which exists for $U(1)$ gauge groups. 
Here, the R-symmetry of the theory forbids the superpotential interactions for the $A$ field non-minimally coupled to the $G^{mn}$ tensor. The spontaneous breaking of supersymmetry during inflation may introduce interactions however these will be generated radiatively and should not affect the tree level D-term inflationary potential. The D-term potential domination together with the enhanced friction features strongly motivates the study of this higher derivative theory to inflationary applications.

\section{Examples of higher derivative D-term inflation and constraints from the CMB}

In our context the gauge kinetic function $\text{Re}f(A)$ is arbitrary and in principle contains non-renormalizable terms. In most of the models, again, one finds that the $|A|$ is of order $M_P$ or larger, a fact that makes the non-renormalizable terms difficult to control similarly to the higher order terms in the $K$ and $W$ potential. However, here it can be $|A|\ll 1$ and inflation can be realized thanks to the nonminimal derivative (also called kinetic) coupling.

We will attempt to capture some of the characteristics of the nonminimal kinetic coupling in inflationary applications by considering some representative examples of  inflationary potentials. We will concentrate on {\itshape single field inflation} models where one of the two fields is heavy enough and stabilized in the vacuum.  
In a FLRW background, neglecting spatial gradients, the Friedmann equation and the equation of motion for the $\phi$  field are 
\begin{equation} \label{mod}
H^2=\frac{1}{3M^2_P}\left[\frac{\dot{\phi}^2}{2}\left(1+9M^{-2}_*H^2\right)+V(\phi) \right], \quad\quad \partial_t\left[a^3\dot{\phi}\left(1+3M^{-2}_*H^2 \right) \right]=-a^3 V_\phi\,.
\end{equation}

According to the equations (\ref{mod}) and for $H M^{-1}_*\gg 1$ the slow-roll parameters of General Relativity (GR) $\epsilon \equiv  {M^2_P} ({V'}/{V})^2/2$ and $\eta\equiv M^2_P V''/V$ are recast into 
\begin{equation}
\tilde{\epsilon} \approx \frac{\epsilon}{3H^2 M^{-2}_*} \,\, , \quad \quad \tilde{\eta} \approx \frac{\eta}{3H^2 M^{-2}_*}\,.
\end{equation}
The requirement $\tilde{\epsilon}, |\tilde{\eta}|<1$ yields that the field space region where slow-roll inflation is realized is rather increased. We will illustrate this below by considering different forms for the gauge kinetic function and thereby various types of potentials \cite{Dalianis:2014sqa,Dalianis:2014nwa}. 

Let us assume  that 
\begin{equation} \label{f1}
f(A) = \frac{\xi^2}{2V_0}\sum_n \lambda_n\left(\frac{ A}{M_P}\right)^n , 
\end{equation}
where $\lambda_n$ are real coefficients and we constrain the field to sub-Planckian values $|A| \ll M_P$. 
The scalar potential reads ${\cal V}(\phi, \beta) = V_0\left(1-\lambda_1{\phi}/{M_P}- (\lambda_2 -\lambda^2_1)\phi^2/M^2_P +\lambda_2\beta^2/{M^2_P}+...\right)$  where the ellipsis correspond to negligible terms. The above potential includes two scalars that have a non-minimal derivative coupling. For $\lambda^2_1\sim \lambda_2>0$ the $\phi$ field can be light enough and the $\beta$ field can be heavy enough ($\gtrsim H$) and stabilized; hence the appearance of any sub-Planckian strong coupling scale \cite{Germani:2011ua} can be avoided. For $\phi\ll M_P$
 the linear to $\phi$ term dominates and the potential reads  
\begin{equation} \label{hill}
{\cal V} \simeq V_0\left(1-\lambda \frac{\phi}{M_P}\right)\,.
\end{equation}
The slow-roll conditions 
 yield the requirements for inflation $\tilde{\eta}=0$ and $\tilde{\epsilon} \approx M^2_P\lambda^2/(2V_0 M_*^{-2} )<1$. 
The linear model (\ref{hill}) 
yields a spectral index $1-n_s=8\tilde{\epsilon}=4\lambda^2M^2_P M^2_*/V_0$ which is related to the number of e-folds by the expression $1-n_s \approx 4\lambda \Delta\phi/N(\phi_*) M_P$. 
For $N(\phi_*) \sim 50$ 
this model yields a spectral index $n_s \sim 0.04$ for $\Delta \phi \sim M_P/\lambda$.  Hence it is possible to have sublplanckian variations for the inflaton field and couplings $\lambda$ of the order of one or even larger.

If we now assume that the gauge kinetic function is of exponential form then we directly get an exponential type potential:
\begin{equation} \label{fexp}
f(A)= \frac{\xi^2}{2V_0} e^{\lambda A/M_P} \quad \Rightarrow \quad {\cal V} =V_0 \frac{1}{\cos(\lambda \beta/M_P)} e^{-\lambda \phi/M_P}\,.
\end{equation}
The form of the function $1/\cos x$ suggests that the $\beta$-dependent part of the potential will be stabilized with large enough mass
to values  $\langle{1/\cos}(\lambda \beta/ M_P) \rangle = 1 $ and the $\phi$ will be the inflating field.  Slow-roll inflation takes place for $\tilde{\eta}=2\tilde{\epsilon}=\lambda^2/(3H^2 M_*^{-2}) <1$ which corresponds to inflaton field values $\phi< M_P/{\lambda} \ln \left({ V_0/ M_*^{2}}{M^2_P \lambda^2}\right)$. The interesting feature of this theory is that inflation naturally ends,  see Ref. \cite{Dalianis:2014nwa} for further analysis. 
For the  exponential potential (\ref{fexp}) 
one finds $\tilde{\eta}=2\tilde{\epsilon}$, $1-n_s=8\tilde{\epsilon}-2\tilde{\eta}=4\tilde{\epsilon}$ and $1-n_s\sim 2/N(\phi_*)$. 
Hence, it predicts a tensor-to-scalar ratio $r=0.16$ for $n_s=0.04$.

An inverse power law potential can be obtained if we consider monomial gauge kinematic function:
\begin{equation} \label{finv}
f(A)= \frac{\xi^2}{2V_0}\, \lambda \frac{A^n}{M^n_P}\quad \Rightarrow \quad {\cal V} =V_0 \frac{1}{\lambda} \frac{M^n_P}{\phi^n}\,,
\end{equation}
where we used the relation Re$\{A^n\}=\phi^n \cos(n\theta)/ (\cos\theta)^n$. We see that the field $\theta$, the phase of the complex field $A=\rho e^{i\theta}$, is stabilized with large enough mass at $\theta =\kappa \pi$ and so Re$\{A^n\}=\phi^n$. 
Slow-roll inflation takes place for $\tilde{\epsilon}=(M^2_Pn^2)/(2\phi^2\times 3H^2 M_*^{-2})<1$, $\tilde{\eta}=2(1+n^{-1})\tilde{\epsilon}<1$ which corresponds to inflaton field values $\phi^{n-2}<2 M^{n-4}_P {V_0 }/{M_*^{2}\lambda n^2}$. 
In the case of the  inverse power-law models (\ref{finv}) the spectral index is given by the modified  expression $1-n_s=4 \tilde{\epsilon}\, (1-n^{-1})$ and it is in tension with the Planck data.

For the above types of potentials an inflationary phase can be realized for a wider range of parameters. For the linear and the exponential, in GR limit, inflation 
is impossible for $\lambda \geq {\cal O}(1)$. It has to be $\lambda<{\cal O}(1)$ which implies, after absorbing $\lambda$ to the mass scale, that the field $\phi$ has to be suppressed by a super-Planckian value. Here thanks to the kinetic coupling inflation is possible even for $\lambda \gtrsim 1$ and for sub-Planckian excursions for the (non-canonical) inflaton field $\phi$. 
The kinetic coupling operates like an enhanced friction and inflation takes place more generically than in the GR limit.

\section*{Acknowledgments}

It is a  pleasure to thank A. Kehagias and C. Germani for our collaborations and discussions. 
This work was supported by the Grant agency of the Czech republic under the grant P201/12/G028.

\end{document}